\documentclass{article}

\usepackage[utf8x]{inputenc}
\usepackage[margin=1.2in]{geometry}
\usepackage{graphicx}
\usepackage{booktabs}
\usepackage{natbib}
\usepackage{ulem}
\usepackage{authblk}
\usepackage{hyperref}
\usepackage[hyphenbreaks]{breakurl}

\graphicspath{{./images/}}
\title{The reported durations of GOES Soft X-Ray flares in different solar cycles}

\author[1]{Bill Swalwell\thanks{Email: bswalwell@uclan.ac.uk}} 
\author[1]{Silvia Dalla} 
\author[2]{Stephen Kahler} 
\author[2]{Stephen M. White}
\author[3]{Alan Ling}
\author[4]{Rodney Viereck} 
\author[5]{Astrid Veronig}

\affil[1]{Jeremiah Horrocks Institute, University of Central Lancashire, Preston, PR1 2HE, UK}
\affil[2]{Space Vehicles Directorate, AFRL/RVBXD, Bldg 570, 3550 Aberdeen Dr. SE, Kirtland
AFB, NM 87117, USA}
\affil[3]{Atmospheric Environmental Research, 2201 Buena Vista Drive SE, Suite 407, Albuquerque, NM 87106, USA}
\affil[4]{NOAA Space Weather Prediction Center, Boulder, USA}
\affil[5]{Institute of Physics \& Kanzelh\"{o}he Observatory, University of Graz, Graz, Austria}

\date{}

\begin{document}
\maketitle


\begin{abstract}

The Geostationary Orbital Environmental Satellites (GOES) Soft X-ray (SXR) sensors have provided data relating to, 
\textit{inter alia}, the time, intensity and duration of solar flares since the 1970s. The GOES SXR Flare List has become 
the standard reference catalogue for solar flares and is widely used in solar physics research and space weather. We report 
here that in the current version of the list there are significant differences between the mean duration of flares which occurred 
before May 1997 and the mean duration of flares thereafter. Our analysis shows that the reported flare timings
for the pre-May 1997 data were not based on the same criteria as is currently the case.

This finding has serious implications for all those who used flare duration (or fluence, which depends on the chosen 
start and end times) as part of their analysis of pre-May 1997 solar events, or statistical analyses of large samples 
of flares, \textit{e.g.} as part of the assessment of a Solar Energetic Particle forecasting algorithm.

\end{abstract}

\section{Introduction}
\label{sec:introduction}

Solar flares are sudden brightenings across the whole of the electromagnetic spectrum, typically from a small spatial region 
in the Sun's corona. They have been known to occur since the middle of the 19th century (\citet{1859MNRAS..20...13C, 
1859MNRAS..20...15H}). Since 1976 they have been classified according to their peak emissions in the 1 - 8 {\AA} band of the 
X-ray Sensors (XRS) \citet{1994SoPh..154..275G}) carried by a series of Geostationary Orbital Environmental Satellites (GOES). 
X-class flares have a peak soft X-ray (SXR) emission of 
10\(^{-4}\) W/m\(^{2}\) or higher; M-class flares a peak SXR emission between 10\(^{-5}\) and 10\(^{-4}\) W/m\(^{2}\); 
C, B, and A-class flares are similarly defined (\citet{2000eaa..bookE2285C}).

Flare duration has been an important parameter for those involved in the field of solar physics for decades. For many 
years flares have been grouped into two types: ``gradual'' or ``long duration'', and
``impulsive''. Gradual flares remain within 10\% of their peak intensity for more than 1 hour, whereas impulsive flares return to 
below that threshold within 1 hour (\textit{e.g.} \citet{1986ApJ...301..448C, 1992ApJ...391..370K}). This classification has formed 
the basis of a large body of work. 

Furthermore, flare duration and fluence have been known to be a significant parameter in relation to the production of Solar Energetic 
Particles (SEPs) within a space weather forecasting environment, as will be shown in Section~\ref{sec:discussion}. 
Evaluation of SEP forecasting algorithms over long time ranges requires a consistent flare duration dataset.

Since 1976 the GOES SXR Flare List has become the standard solar flare catalogue. The list may be accessed through 
a number of different sources: \textit{e.g.} directly from NOAA's National Centers for Environmental Information 
website, through the Heliophysics Integrated Observatory (``Helio'') website
(\citet{2012ASPC..461..255A}), and by using routines in both SolarSoft SSWIDL (\citet{Freeland1998}) and SunPy 
(\citet{1749-4699-8-1-014009}). In SolarSoft the list is retrieved by calling the routine ``get\_gev'' with 
a specified start and end time. In SunPy the relevant routine is ``sunpy.instr.goes.get\_goes\_event\_list''. Relevant 
URLs for Helio, SolarSoft, and SunPy are given in the Acknowledgements Section.

A significant difference in the reported mean duration of X-class flares between a time range incorporating Solar Cycles 21 and 
22, and one incorporating Cycles 23 and 24 was noted by \citet{2017SoPh..292..173S}. Those authors did not seek to explain the 
discrepancy.

In this work differences between mean flare duration as reported by the GOES SXR Flare List in different solar cycles are analysed. 
Flare data are now available for four full solar cycles. The GOES SXR Flare List which is used in our analysis below was 
obtained from the Helio website
(\citet{2012ASPC..461..255A}), but these results have been independently confirmed 
using other files on the National Geophysical Data Center website.

\section{Data Analysis}
\label{sec:flare_duration}

The start time of a GOES SXR flare, as currently defined by NOAA, is the time when 4 consecutive values in the 1-minute 1-8 {\AA} data meet 
all 3 of the following conditions:

\begin{itemize}
 \item All 4 values are above the B1 threshold
 \item All 4 values are strictly increasing
 \item The last value is greater than 1.4 times the value which occurred 3 minutes earlier
\end{itemize}

The peak time of the flare is when the SXR flux reaches its maximum (and it is the value of the SXR flux at this time which 
defines the class of the flare). The flare end time is defined as the time when the flux reading returns to 1/2 the 'peak', 
where the peak is the flux at maximum minus
the flux value at the start of the event. Here we take flare 
duration to be the total time between the reported flare start time and flare end time; ``rise time'' is the time between 
flare start time and the time of flare maximum; and ``decay time'' is the time between the time of flare maximum and the 
flare end time. At the time of writing, events with fast rise times are derived automatically by an algorithm processing the 
SXR data, whereas those with slow rise times are recorded manually.  

\begin{figure}
 \centerline{\includegraphics[width=.8\textwidth]{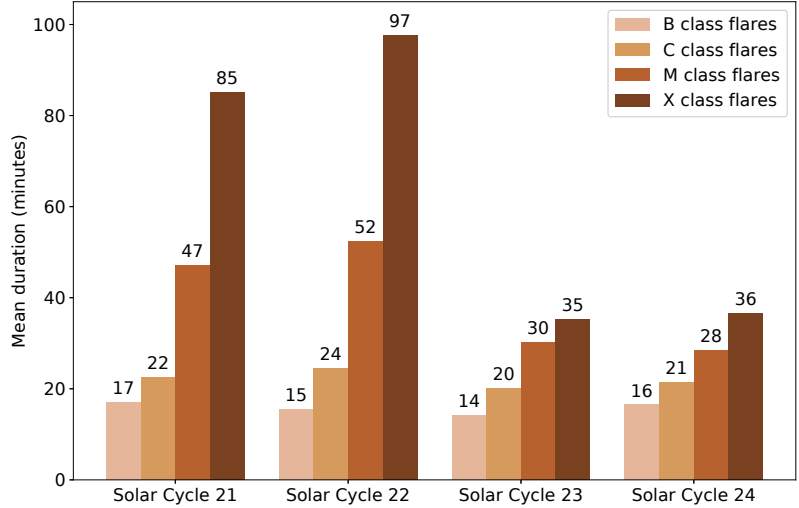}}
 \caption{Mean reported duration (in minutes) of flares of different classes in each of the last 4 solar cycles as derived from 
 the GOES SXR list. From left to right bars represent B-class flares, then C-class, M-class, and X-class.}
 \label{fig:mean_duration}
\end{figure}

Figure~\ref{fig:mean_duration} is a bar plot showing the mean duration (in minutes) of flares of different classes in each 
of the last 4 solar cycles derived from the GOES SXR Flare List: in this work Solar Cycle 21 is taken to have started 
on 1 January 1976, Cycle 22 on 1 January 1986, 
Cycle 23 on 1 January 1996, and Cycle 24 on 1 January 2008. From left to right the bars represent B-class flares, then C-class, 
M-class, and X-class. It is readily apparent that the mean reported duration of both M and X-class flares in Solar Cycles 21 
and 22 is much longer than in Cycles 23 and 24. 

As the difference in mean duration is most apparent for flares of a higher class, we concentrate on X-class flares. We plotted 
the 1-minute time-averaged SXR data for each reported X-class flare from 1 January 1986 onwards (as the NOAA website does not 
publish such data for earlier periods). Table~\ref{tab:sc22_flares} shows the reported timings of a representative sample of 4 
X-class flares in the GOES SXR List which occurred in Solar Cycle 22. Column 1 gives the flare class, column 2 the date of the 
event, and columns 3, 4, and 5 the reported start time, peak time, and end time of each flare.

 \begin{table}
\centering
\caption{The reported timings of a sample of 4 X-class flares which occurred during Solar Cycle 22. Column 1 gives the flare class, 
column 2 the date of the event, and columns 3, 4, and 5 the reported start time, peak time, and end time of each flare.}
 \begin{tabular}{l c c c c}
 \toprule
 Flare class & Date & Reported start & Reported peak & Reported end \\ 
 \midrule
 X1.6 & 1988-06-23 & 08:56 & 09:27 & 10:03 \\
 X2.4 & 1988-06-24 & 16:03 & 16:48 & 16:54 \\
 X1.1 & 1989-01-07 & 04:12 & 04:36 & 04:44 \\
 X2.3 & 1989-01-13 & 08:29 & 10:18 & 10:45 \\
 \bottomrule
 \end{tabular}
\label{tab:sc22_flares}
\end{table}

\begin{figure}
 \centerline{\includegraphics[width=.9\textwidth]{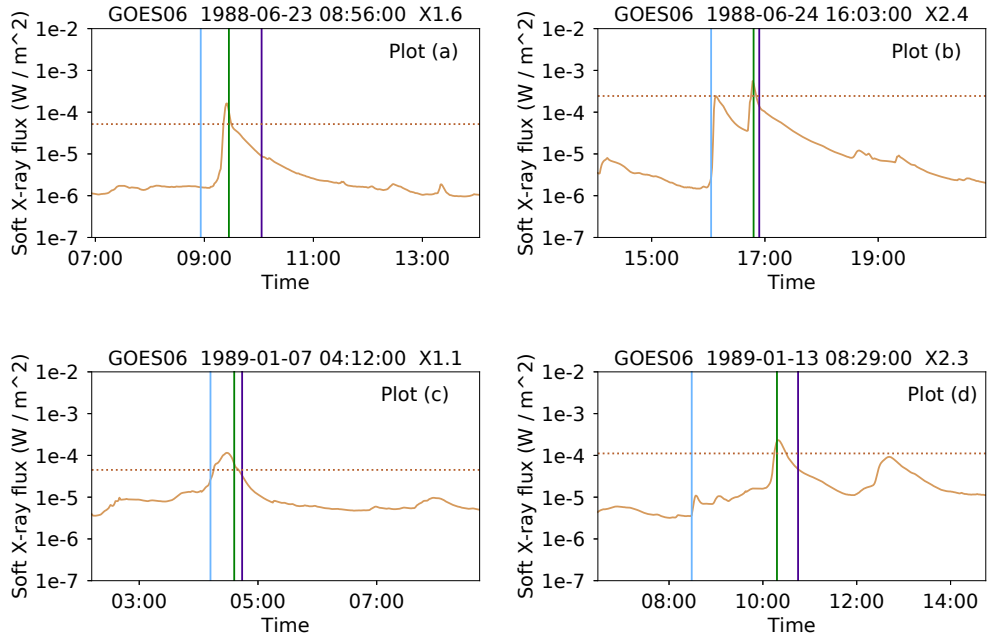}}
 \caption{Soft X-ray flux for a sample of 4 X-class flares in Solar Cycle 22. Time is shown on the x-axis, and the 1-minute 
time-averaged soft X-ray flux in W/m\(^{2}\) on the y-axis. On each plot the vertical light-blue line is drawn at the flare's 
reported start time; a vertical green line at its reported peak; and a vertical purple line at its reported end. The horizontal 
dotted brown line is drawn at half the peak of the SXR flux (as defined by NOAA).}
 \label{fig:sc_22_x_flares}
\end{figure}

Figure~\ref{fig:sc_22_x_flares} shows plots of the 1-minute 
time-averaged SXR downloaded from the NOAA website
for each of this sample of 4 flares. Time is plotted 
on the x-axis: the starting point for each plot was 2 hours prior to the reported start time of the flare, and the end point was 
6 hours after its reported end. On the y-axis is plotted the 1-8 {\AA} 1-minute time-averaged SXR flux in W/m\(^{2}\).  

On each plot a light-blue vertical line is drawn at the flare's start time as reported in the catalogue; a vertical green line 
at its reported peak; and a vertical purple line at its reported end. The horizontal dotted brown line is drawn at half the peak 
of the SXR flux as previously defined(which represents the end of the flare according to the NOAA criteria). The name of the GOES 
spacecraft carrying the SXR sensor is specified at the top of each plot, as is the reported start time of the flare and its 
reported class. 

For the flare shown in plot (a) it can be seen that the reported start time is several minutes earlier than the actual start of the 
rise in SXR flux; the reported peak is slightly different from the actual peak; and the reported end of the flare is many minutes later 
than it ought to be according to the NOAA definition. Plot (b) shows the SXR flux of an X2.4 flare which occurred the day after the 
flare shown in plot (a). Here, there were 2 X-class flares in quick succession, but only 1 is reported, and the times of the 2 flares have 
been combined - the reported start of the flare is for the first of the 2 events, but the reported peak and end are for the second flare. 
For the flare shown in plot (c) reported start and end times are slightly awry, and the reported peak is some time later than the peak in 
SXR flux; and in plot (d) both reported start and end times do not appear to accord with the NOAA definition.

To illustrate that the qualitative behaviour seen in Figure~\ref{fig:sc_22_x_flares} is ubiquitous, we considered flares of class \(\geq\) 
M5 and developed a method of calculating rise and decay times directly from the SXR flux time series. To obtain the flare start time we took 
the time of the peak as originally reported and looked back to find the time when the SXR flux was either 5\% of 
the peak flux, or where the slope (\textit{i.e.} the derivative) of a highly smoothed long-channel light curve reached 5\% of the peak slope, 
whichever time was the later. The value of 5\% was chosen so as to exclude pre-flare heating, and to ensure that if there had been another 
peak prior to the flare of interest the start time would fall between the two flares. To find the flare end time we looked forward from the 
originally reported peak time to find the time where the SXR flux fell to 50\% of the peak value. Whilst the method was surprisingly accurate 
in finding flare start time, in a small number 
of cases the timing of the start of the flare was adjusted manually based upon inspection of the data. 

\begin{figure}
\centerline{\includegraphics[width=.8\textwidth]{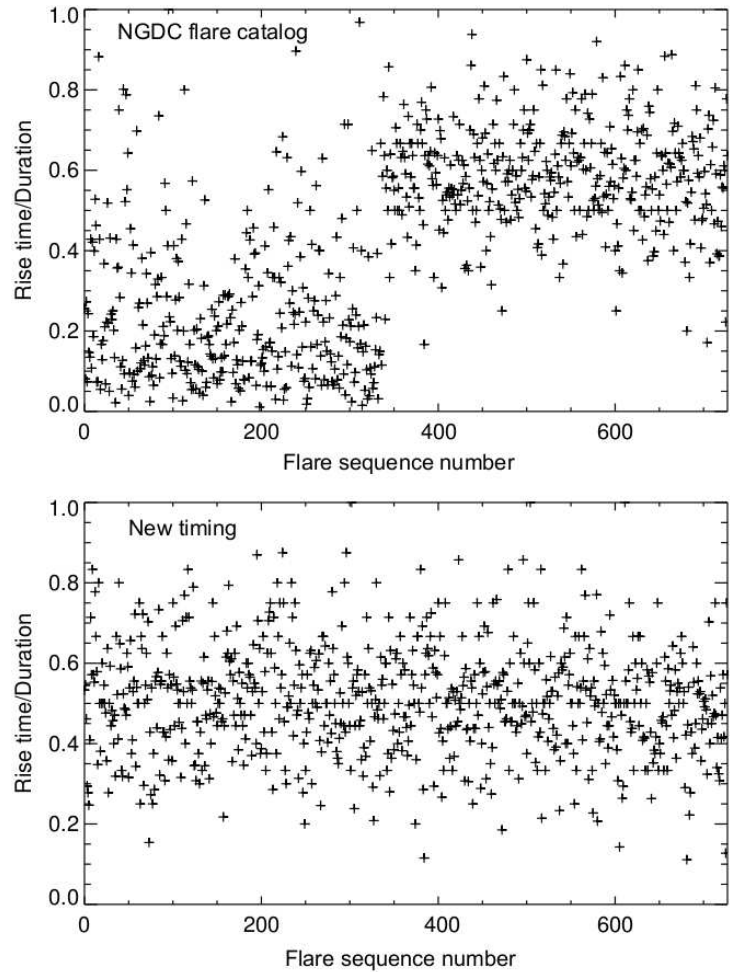}}
\caption{Plots of the ratio of flare rise time to total flare duration for flares of class \(\geq\) M5 between 1986 and 2015. In the top plot 
the ratio is derived using timings from the GOES SXR Flare List, whereas the ratio for the bottom plot is derived from our timings.}
\label{fig:stephens_plots}
\end{figure}

Figure~\ref{fig:stephens_plots} compares flare rise times as a fraction of total flare duration for flares greater than class M5 between 1986 
and 2015. The ratio of rise time to duration appears on the y-axis, and flare sequence number on the x-axis. The top 
plot of Figure~\ref{fig:stephens_plots} shows the original timings as reported in the GOES SXR Flare List: the ratio is centred around 
0.19 (median) for flares which occurred prior to 1997, but the ratio changes to be centred around 0.58 (median) after 1997. The bottom plot 
of the same Figure shows the same ratio but in this case based upon our timings, and for both 
pre and post 1997 flares the ratio remains centred at a median value of 0.50.

It is clear from Figure~\ref{fig:stephens_plots} that a significant change occurred in 1997. With a view to discovering when in 1997 this
happened, we examined plots similar to those shown in Figure~\ref{fig:sc_22_x_flares} for the more frequent M-class flares. It is apparent 
that the reported flare timings up to and including the M1.9 flare on 1 April 1997 do not accord with the NOAA definition, whereas the timings 
of the next M-class flare (which was an M1.3 flare on 21 May 1997) do accord with that definition. The change in the way that flare timings are 
reported occurred within that nearly two month period.

\begin{figure}
\centerline{\includegraphics[width=.8\textwidth]{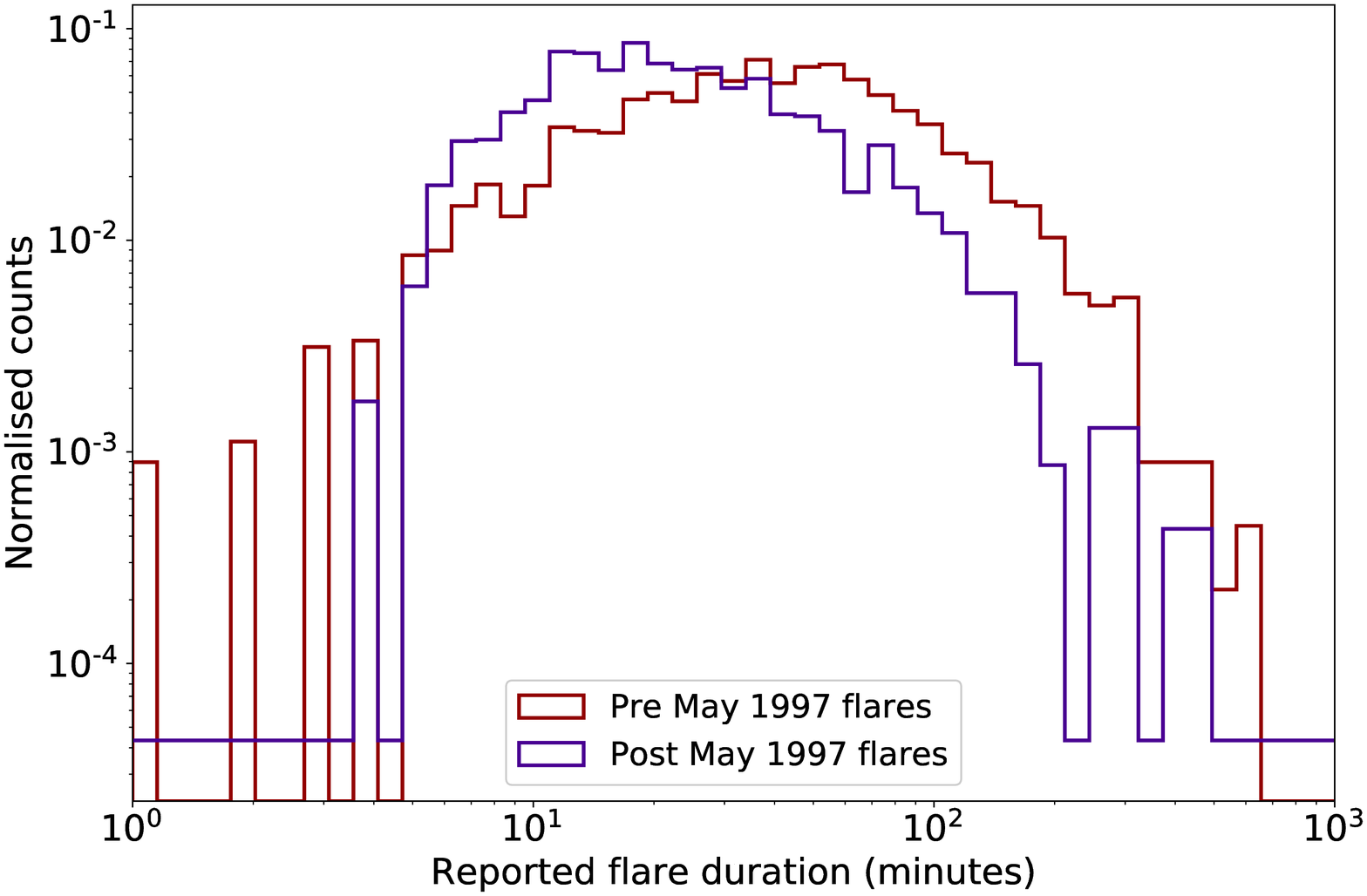}}
\caption{Distribution of reported flare durations for M and X-class flares in the GOES SXR Flare List 
for the time range prior to May 1997 (brown line) and after May 1997 (purple line). Flare counts are normalised 
to the overall number of flares in each time range.}
\label{fig:hist_duration}
\end{figure}

We also considered the distribution of flare duration shown in Figure~\ref{fig:hist_duration} considering M and X-class flares only.
The distribution for the period prior to May 1997 (brown line) is compared with that 
post May 1997 (purple line). It is readily apparent that there was a greater proportion of large flares which were reported to 
have a duration of less than about 30 minutes post May 1997. Conversely there was a greater proportion of large flares 
reported to last longer than about 30 minutes prior to May 1997.

\section{Discussion}
\label{sec:discussion}

Our analysis of the GOES SXR Flare List shows that there are clear systematic differences in mean flare duration between a time range 
including Solar Cycles 21 and 22, and one including Solar Cycles 23 and 24. The effect is particularly clear for X and M-class flares: 
the mean duration of X-class flares in Cycles 21 and 22 was respectively 2.4 and 2.7 times as long as that for Cycle 23; for M-class 
flares the mean duration for Cycles 21 and 22 was respectively 1.6 and 1.7 times as long as that for Cycle 23. 

\citet{2002A&A...382.1070V} reported that prior to 1997 the reported SXR flare times were taken from the associated H\(\alpha\) event. 
These timings were originally reported in the Solar-Geophysical Data Reports (commonly called the ``Yellow Books'') and which are now 
mostly available online. The table headed ``GOES Solar X-ray Flares'' in those 
books often has an ``Editor's Note'' at the bottom which reads ``\textit{Please note that whenever optical flares are given, the times 
given are times of the optical flares and not the times of the X-ray flares}''. Our analysis indicates that this is the case for most, 
especially large, flares but we have not checked all the data manually. This information, however, is not propagated within the 
tools such as Helio, SolarSoft, or Sunpy. 

H\(\alpha\) flare duration is defined visually, \textit{i.e.} how long 
the flare can be seen, and the timings given in the Yellow Books are based upon reports from many different observing stations. It is 
therefore entirely unsurprising that these times do not in general correspond with the definition of flare timings published by NOAA.
It seems, therefore, that the differences reported here stem from a change of use of H\(\alpha\) timings to timings based upon SXR 
flux as measured by the GOES X-ray Sensors. Whatever the cause, pre May 1997 flare timings are not directly comparable with post May 
1997 flare timings.

This finding can have serious implications for some statistical studies that used the GOES X-ray flare listings prior to May 1997.  However, 
we have to be careful to distinguish those works that used the flare listings for only the correct peak X-ray fluxes (\textit{e.g.} 
\citet{2004SpWea...2.2002G, BELOV2009467}) and not for times or fluences. Further, many authors used the pre-1998 GOES XRS flux-time 
profiles to determine independently their own flare times and fluences (\textit{e.g.} \citet{1986ApJ...301..448C, 2008SpWea...6.1001B, 
2009SpWea...7.4008L, 2014JGRA..119.9383J, 2015SoPh..290..819T, 2016JSWSC...6A..42P}) or used those independent lists for further analyses 
(\textit{e.g.} \citet{doi:10.1002/2015SW001222, SWE:SWE20206}).  Finally, there have been many SEP event studies based on X-ray 
flare reports together with Coronal Mass Ejections (CMEs) from the SOHO/LASCO catalog listings (\textit{e.g.} \citet{2013SoPh..282..579M, 2014JGRA..119.9456P, 
2015SoPh..290..841D, 2017Ge&Ae..57..727B}).  Those CME listings began in January 1996, so there is an overlap of CME reports and GOES SXR 
flare listings from that time to May 1997.  During that period of low solar activity there were only seven \(>\)M1 flares, two \(>\)M3 flares, 
and no NOAA \(>\)10 pfu at \(>\)10 MeV SEP events.  The impact of the incorrect flare listings on those SEP studies and on flare-CME comparisons 
(\textit{e.g.} \citet{2009IAUS..257..233Y}) should therefore be minimal.

We know of significant impacts to two (involving current authors) recent reports on SEP events.  In their validation of the Proton Prediction 
System (PPS) \citet{refId0} calculated X-ray flare fluences from 1986 to 2014 as the product of the flare rise times (onset to peak) and the 
peak fluxes obtained from the NOAA listings.  Of their 716 \(>\)M5 X-ray flare candidates, 344 were before May 1997, as were 26 of their 67 SEP events.  
The incorrectly reported flare rise times in the listings before May 1997 (shown in the top panel of Figure 3) would suggest that \citet{refId0} used 
inaccurate X-ray fluences, which would have affected the forecasting of SEP events with PPS for that time. The PPS validation with three groups 
of 8800 MHz bursts in their work was independent of the X-ray fluences and remains valid.  

In the second impacted report \citet{2017SoPh..292..173S}
defined two algorithms to forecast \(>\)40 MeV SEP events. Their second algorithm using X-class flares to forecast SEP events was tested over 
two time ranges: 1996 to 2013 and 1980 to 2013.  While that algorithm was based only on flare intensities, they also displayed the flare 
durations in their Figure 11, which shows much longer X-class flare durations for the two solar cycles before 1997 than for the two following 
cycles. This discrepancy led to the current investigation of the NOAA X-ray flare reports.  Fortunately, it does not affect their validations 
of the two forecasting algorithms.

In the next year, NOAA will be reprocessing many years of XRS data and publishing it in the same format as that of GOES-16 and subsequent 
satellites. This reprocessing will result in a consistent flare event list with start, peak, and end times, as well as integrated flux. The 
processing will also include a number of fixes and include both corrected fluxes and a NOAA flare index consistent with the current flare 
values.

\section*{Acknowledgements}

The GOES SXR flare list may be accessed through the NOAA website at 
\url{https://www.ngdc.noaa.gov/stp/space-weather/solar-data/solar-features/solar-flares/x-rays/goes/xrs/}, 
through the Helio website at
\url{http://www.helio-vo.eu/} and via routines in SSWIDL (see \textit{e.g.} 
\url{https://hesperia.gsfc.nasa.gov/rhessidatacenter/complementary\_data/goes.html}) and SunPy (see \textit {e.g.} 
\url{http://docs.sunpy.org/en/v0.9.0/\_modules/sunpy/instr/goes.html\#get\_goes\_event\_list}). The time-averaged SXR data may be 
accessed through \url{https://www.ngdc.noaa.gov/stp/satellite/goes/dataaccess.html}. The Yellow Books may be downloaded at 
\url{https://www.ngdc.noaa.gov/stp/solar/sgd.html}.

SD acknowledges support from the Leverhulme Trust (grant RPG-2015-094). SK was funded by AFOSR Task 18RVCOR122.  AL was supported by 
AFRL contract FA9453-12-C-0231. AV acknowledges the support by the Austrian Science Fund (FWF): P27292-N20.

We should also like to thank the two anonymous referees for their helpful comments.


\end{document}